# Off-axis optical trapping and transverse spinning of metallic microparticles with a linearly polarized Gaussian beam

Yansheng Liang[1], Zhaojun Wang[1], Ming Lei[1*]

[1] *Shaanxi Key Laboratory of Quantum Information and Quantum Optoelectronic Devices, School of Science, Xi'an Jiaotong University, Xi'an 710049, China*

* *ming.lei@mail.xjtu.edu.cn*

**Abstract:** Optical trapping of metallic microparticles remains a big challenge because of the strong scattering and absorption of light by the particles. In the paper, we report a new mechanism for stable trapping of metallic microparticles by using a tightly focused linearly polarized Gaussian spot. We theoretically and experimentally demonstrated that metallic microparticles were confined off the optical axis by such a trap. In the meanwhile, transverse spinning motion occurred as a consequence of the asymmetric force field acting on the particle by the trap. The off-axis trapping and transverse spinning of metallic microparticles provide new manners for the manipulation of metallic microparticles. The works reported in this paper are also of significance for a better understanding of the mechanical interaction between light and metallic particles.

Key Words: optical trapping, optical spinning, metallic microparticles, off-axis trapping, transverse spinning

## References

## 1. Introduction

Metallic particles are drawing increasing attentions because of their size- and shape-dependent optical properties [1]. They have served as remarkable tools in various applications, such as drug carriers in drug delivery[2], label-free biosensors in bio-sensing [3, 4], and optical nano-antennas in optical communications [5, 6]. They are also exploited as remotely, optically controlled micro/nanoscopic sources of heat for bubble formation[7], surface ejection[8], and cancer therapy[9, 10]. Widely growing applications imply that the optical trapping of metallic particles would be of great interest to a variety of scientific communities.

During the past two decades, theories and experiments focusing on optical trapping of metallic particles have been vigorously studied. Metallic nanoparticles are proved as better candidates for optical trapping due to the higher polarizability compared with dielectric nanoparticles with equivalent size [11-17]. The gradient force is much larger for a metallic nanoparticle than a dielectric one, while the scattering force is negligible. Consequently, higher trapping efficiency can be achieved even by using a bright spot, such as a tightly focused Gaussian spot. However, as the particle size increases, the scattering force arising from scattering and absorption of light increases dramatically and dominants, making large metallic particles become poor candidates for optical trapping. Efforts have been made to find the solutions, especially for three-dimensional (3D) trapping [18-22]. The simple means to confine a metallic microparticle is using a structured hollow trap, for example the vortex beam [23, 24]. The scattering force pushes the metallic particle away from the center of a bright spot. Therefore, hollow trap can be utilized to confine the particle to the dark center. The non-ignorable drawback is that the particle will be trapped in the transverse plane, and forces acted by a wall or gravity are needed to cancel the axial scattering force to achieve 3D confinement. Gaussian beam has also been demonstrated for two-dimensional trapping of gold Mie particles in the transverse plane [18, 19]. Further theoretical analysis performed by Gu et al. showed that the trapping efficiency of metallic Mie particles could be enhanced by the use of a centrally obstructed Gaussian beam [20, 21]. While the theoretical analysis works, the experimental demonstration of 3D trapping of metallic microparticles (diameter ≥ wavelength) is still an ongoing task [22].

In this paper, we report a new mechanism for stable trapping of metallic microparticles. We demonstrate the off-axis confinement of metallic microparticles by the use of a Gaussian beam with numerical analysis and experimental results. While the metallic microparticle is confined at the off-axis equilibrium position, it will do transverse spinning motion as a consequence of the asymmetric force field that the particle experiences. This phenomenon is intrinsically different from the optical spinning utilizing the transfer of photon spin angular momentum [25, 26], or the specially designed shape of particle that leads to spinning torque [27]. The off-axis transverse spinning of metallic particles relies on the asymmetric light field introducing asymmetric force field, which has never been reported by the previous works. The transversely spun particles are expected to serve as a new kind of optically driven rotators that can be applied to biology or hydrodynamics [28].

## 2. Methods and materials

The experiments were conducted with an inverted single-beam optical tweezers setup, as illustrated in Fig. 1(a). The trapping laser running at a wavelength of 1.064μm has a maximum output power of 5W. The collimated beam is directed into a high numerical-aperture oil-immersion objective (100×/NA1.45, Nikon Inc., Japan) by a dichroic mirror. A CMoS camera (GS3-U3-41C6M-C, Point Grey Research Inc., USA) with resolution of 2048×2040 pixels and pixel pitch of 5.5 μm is employed to monitor and record the manipulation process. A white light source focused by a condenser is used to illuminate the samples for direct wide-field imaging. We use gold particles (Thermo Fisher Scientific Inc., USA) with the diameter ranging from 1.0 to 5.0 μm as metallic candidates for trapping.

Figures 1(b-c) present the principle of off-axis optical trapping of metallic microparticles by the use of Gaussian beams. For large metallic particles, scattering force dominates, propelling the particles away from the intensity maximum. However, we find that metallic particles can be trapped at the edge of the focal spot (Fig. 1(b)). At this position, the longitudinal scattering force is balanced by the gravity of the particle, and the transverse scattering force is too weak to push the particle away. Indeed, the particle will be pulled back if it moves away from the beam center by a vortex-like force field, which will be discussed below. The off-axis trapped particle will bear asymmetric force field at its two ends (Figs. 1(c, d)). For the case given in Fig. 1(c), the force acting on the particle at the right end is larger than that at the other end, making the particle roll in the anticlockwise direction around the transverse axis. In comparison, the metallic microparticle trapped at the position given in Fig. 1(d) will roll clockwise around the transverse axis. This kind of rolling also can be referred to *transverse spinning* as the particle spins around the transverse axis.

## 3. Results

### 3.1 Simulation results

The optical forces exerting on a metallic particle by an optical trap include absorption force, scattering force, and gradient force. For particles with a diameter larger than half wavelength of trapping light, the extinction increases dramatically with the particle's size [15], and it cannot be neglected. All three kinds of forces should be taken into account for the calculation of the trapping force acting on large metallic microparticles. Here we employ the T-matrix method [29, 30] to calculate the net optical force. We use gold microparticles as trapping candidates whose dielectric constants is $\hat{\varepsilon}=-54+5.9i$ at the wavelength of 1064 nm [31]. In all simulations, we used a y-polarized Gaussian beam.

To verify the trapping performance of a Gaussian beam on gold microparticles, we calculate the two-dimensional forces in the $xz$ and yz planes experienced by a gold particle with a radius of 1.5μm (Fig. 2). The net force field takes into account the gravity of particle, the buoyancy force that the particle experiences, and the optical force acting on the particle by the trap. The trapping power is $P$=10mW. Figures 2(a,d) clearly show that the particle will fail to be trapped on the axis because of the nonzero force. But interestingly, the positions of zero force are found at the positions of $(x_0, z_0) = (\pm 2.62, -0.46)$ μm in the regions 1 and 2 in the xz plane. Moreover, the vector maps in the regions 1 and 2 present two vortex-like vector field with force singularities in the center, but have opposite spiral directions (Fig. 2(b,c)). As a result of the vortex-like force field, the metallic microparticle is expected to be finally trapped at the singular points, i.e., the positions $(x_0, z_0) = (\pm 2.62, -0.46)$ μm. Similarly, zero-force positions can be found in the regions 3 and 4 of the yz plane. However, the vector maps in regions 3 and 4 show no singular points for stable confinement of metallic microparticles (Fig. 2(e,f)).

To confirm that the gold microparticle can be trapped at the positions of $(x_0, z_0) = (\pm 2.62, -0.46)$ μm in the xy plane, we calculate the three-dimensional force curves at the balance positions. As an example, Fig. 3 presents the forces experience by a 1.5-μm-in-radius gold particle at the position of $(x_0, y_0, z_0) = (-2.62, 0, -0.46)$ μm. Results show that, both the longitudinal (Fig. 3(a)) and the transverse force curves (Fig. 3(b,c)) in the vicinity of the balance position have negative slope. The maximal forces are large enough to confine the particle (~0.7pN in the z direction, ~0.26pN in the x direction, and 0.5 pN in the y direction). Therefore, gold particles with a radius of 1.5μm can be stably confined at the position of $(x_0, y_0, z_0) = (-2.62, 0, -0.46)$ μm. Similarly, the particle can be trapped at the position of $(x_0, y_0, z_0) = (2.62, 0, -0.46)$ μm by using such a symmetric Gaussian trap. With these results, we conclude that, by the use of a linearly polarized Gaussian beam, metallic microparticles can be trapped off the axis. This trapping mechanism is quite different from the previously reported on-axis trapping of dielectric particles with a linearly polarized Gaussian beam, in which case there is no vortex-like force field off the axis [20-22].

Compared with the on-axis optical levitation, in which case the metallic microparticle escapes easily from the trap, the off-axis trapping of metallic microparticles tends to be more stable in some sense. When the trapping power is changed, the off-axis trapped gold microparticle will remain stably confined. Although in the meantime, the equilibrium position will change accordingly. For example, for trapping power $P$=10, 50 and 100mW, simulation results show that the equilibrium positions $(x_0, z_0)$ are $(\pm 2.62, -0.46)$, $(\pm 3.75, -0.24)$ and $(\pm 4.33, -0.18)$ μm, respectively. By increasing the trapping power, we found that the particle tends to be pushed away from the beam center in the transverse direction.

Another appealing property besides the off-axis trapping is the vortex-like force field that the off-axis trapped particle experiences. As demonstrated by the above simulation results, the vortex-like force field can confine the metallic microparticles at the singular points. Importantly, it is easy to infer that the vortex-like force field will introduce a spinning torch to the particle, driving it do transverse spinning. To validate this effect, we calculate the spinning torch acting on the particle while it is confined at the position of $x$>0 in the xz plane by a y-polarization Gaussian beam. For a particle with a radius of 1.5μm and trapping power of 100mW, the spinning torch in the x, y and z directions, $(\Gamma_x, \Gamma_y, \Gamma_z)$, are $(7.92\times10^{-17}, 3.94, -2.21\times10^{-16})$ pN·μm, respectively. Compared with the spinning torch in the y-direction, $\Gamma_y$, the components $(\Gamma_x,\Gamma_z)$ in the x- and z- directions are negligible. Consequently, the particle will spin around the y-axis. This phenomenon is quite different from the conventional optical spinning arising from the transfer of spinning angular momentum.

**3.2 Experimental results**

*3.2.1 Off-axis trapping*

As we used a linearly polarized laser as the trapping light source, the metallic microparticle was expected stably confined at two edge positions of the focused Gaussian spot according to the theoretical analysis. Figure 4 presents the time-lapse images of two metallic microparticles ($r$~1.5 μm) stably confined at two edge sides (p1 and p2) of the focal spot (marked by a white dot). When the sample stage moves along the x-direction in the transverse plane, the two particles keep stably confined around the focal spot (Figs. 4(a-d)). In particular, when the stage moves too fast, the particle will be dragged away by the fluid (see the particle at the position of p2 in Fig. 4(c)). But owing to the restoring force acting on the particle by the trap, it will be pulled back to the original position (Fig. 4(d)). When the sample stage moves to the left along the y direction, the two particles still keep stably trapped, but will be dragged to the left side of their original positions by the friction of fluid (Figs. 4(e-h)). To verify that the particles were stably trapped in three dimensions, we translated the sample stage along the z-direction (Figs. 4(i-l)). The two particles stay in the trap during the axial movement of the sample stage. But the distance between the particles increases, and the image gets a little blurred. This means that the equilibrium position has been changed when the particles gets deeper in the solution. It can be explained by decreasing trapping efficiency due to the wavefront distortion caused by the mismatch of the refractive index when using an oil-immersion objective. After the stage moves along the z-direction over a distance of about 10μm (Fig. 4(l)), the stage is translated along the x-direction again. Still, the two particles are stably held in the trap. The time-lapse images given in Fig. 4 demonstrate the off-axis 3D confinement of metallic microparticles by the use of a Gaussian beam (also see Visualization 1).

In the on-axis optical levitation of metallic microparticles, the particles appeared quite unstable as reported by the previous works [22]. In comparison, the off-axis trapped metallic microparticle is expected more stable according to the theoretical analysis, especially for large metallic microparticles. We validated this by investigating the off-axis trapping of metallic microparticles with various trapping power (see Visualization 2). When the output laser power is about 10mW, the metallic particle is confined near the center of the focal spot (Fig. 5(a)). The particle keeps stably trapped when the output power increases, but with increasing distance between the particle and the focal spot center (Figs. 5(b-f)). For output power of $P$= 10, 50, 100, 200, 500 and 1000 mW, the distances are about 1.92, 2.86, 3.19, 3.67, 4.52 and 5.74 μm, respectively. Experimental results show that the off-axis trapping of metallic microparticles can be achieved regardless of trapping power, as long as the power is large enough. Therefore, in some sense the off-axis trapping of metallic microparticles is much more stable than the on-axis levitation.

*3.2.2 Off-axis transverse spinning*

Theoretical analysis has shown that the off-axis trapped metallic microparticle will do the transverse spinning. For clear visualization of this phenomenon, we select an asymmetric metallic particle to be trapped, which is indeed an ensemble consisting of two large ($r$~4μm) particles adhered to each other and several smaller ($r$~1μm) particles on the surface of the two large ones (Fig. 6). Experimental results obtained from the time-lapse video of the off-axis trapped ensemble agree well with the theoretical prediction. The ensemble is trapped at one of the two hot spots. In the beginning, the small particle marked by the yellow arrow lies in the right side of the ensemble (Fig. 6(a)). Then this particle moves above the ensemble (Fig. 6(b)) and further to left side (Fig. 6(c)). The ensemble is transversely spinning in the anticlockwise direction from the view as the dashed arrow indicated. At *t*=60~80ms (Figs. 6(d-f)), the light reflected by the surface of the trapped particles (marked by the dashed squares and circles) also indicates a transverse spinning motion of the ensemble, as the intensity increases gradually when the small particle is getting to the top of the ensemble. The spinning speed measured about 6.3 Hz (see Visualization 3). When the ensemble is trapped at the other hot spot, it will do transverse spinning as well, but in the opposite direction (see Visualization 4). In this case, the spinning rate measured about 3.4 Hz. The difference of the spinning rate at the two spots relies on the beam profile, the shape of the ensemble and the trapping position, or other factors.

## 4. Discussion and Conclusion

In summary, we have reported a novel mechanism for optical trapping of metallic microparticles. We demonstrated the off-axis optical trapping and transverse spinning of metallic microparticles using a linearly polarized Gaussian beam with theoretical analysis (Fig. 3) and experimental measurement (Figs. 4 and 5). The off-axis trapping of metallic microparticles is much more stable than the on-axis levitation, along with changing transverse trapping position dependent on the trapping laser power. The off-axis optical trapping of metallic microparticles provides a new route for manipulating of metallic microparticles.

By the use of a linearly polarized Gaussian beam, metallic particles are demonstrated to be confined at the two hot spots shown in Fig. 4. When the particles are placed at other positions near these two hot spots, they will be attracted to the hot spots. This feature can be exploited to assemble metallic microparticles along an arc as the dashed rings show in Fig. 7 (also see Visualization 5). Experimental results show that the size of metallic microparticles has a significant impact on trapping stability. Large particles will be stably trapped off the axis as previous results demonstrated, whereas small particles seem much less stable. This is because small metallic microparticles will experience violent Brownian motion, thus making itself jitter in the trap due to the vortex-like force field (see Visualization 5).

While trapped off the axis with a Gaussian trap, the metallic microparticle was demonstrated to do transverse spinning simultaneously (Fig. 6), which is caused by the asymmetric force field that the particle bears when it is not at the center of the Gaussian spot. This is intrinsically different from the optical spinning utilizing the transfer of photon spin angular momentum. The transverse spun metallic microparticle can serve as a new source of controllable micro-flow in thermodynamics, and a new kind of spanner in molecular biology. Our works on the mechanism of off-axis trapping and transverse spinning of metallic microparticles in this paper also provide a better understanding of interactions between light and metallic particles.

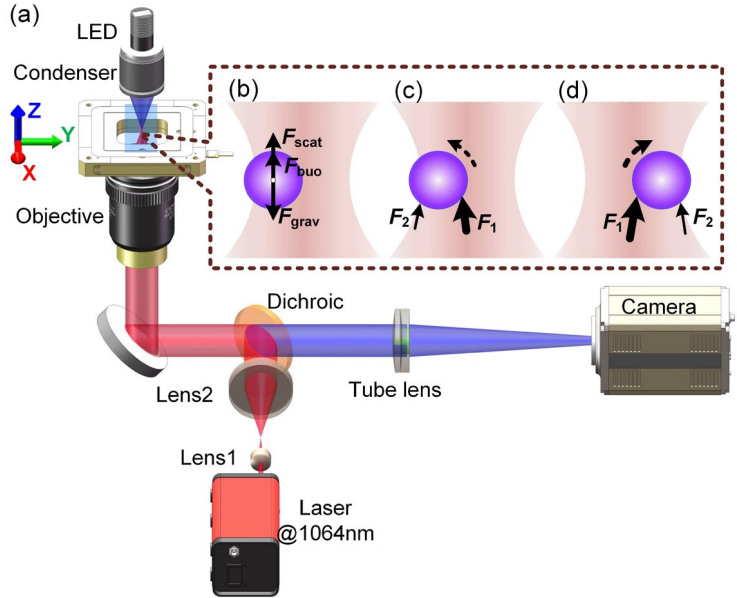

Fig. 1. Sketch for the off-axis trapping of metallic microparticles with an inverted optical tweezers setup. (a) Sketch for an inverted optical tweezers setup. (b) The principle of off-axis optical trapping of metallic microparticles. The scattering force is balanced by the gravity of particle. (c, d) The principle of transverse spinning of metallic microparticles. The asymmetric force exerting on the particle will drive the particle to spin transversely.

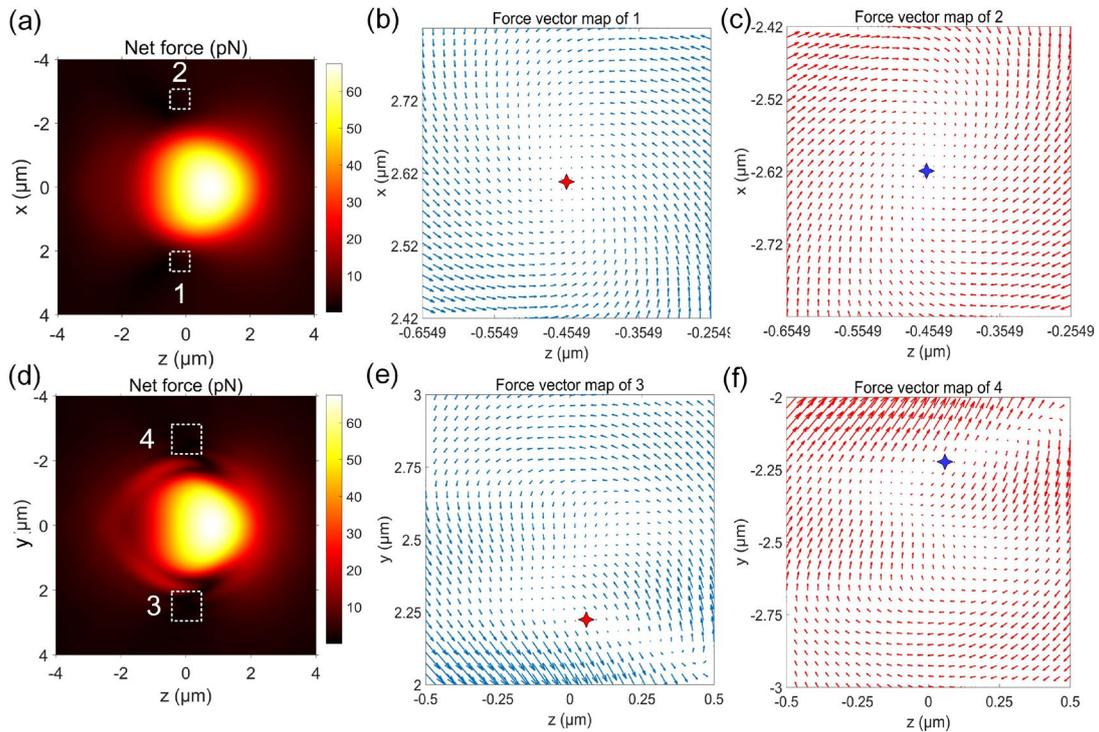

Fig. 2. Force fields of a 1.5-μm-in-radius gold particle trapped with trapping power of 10 mW. (a) Force map. (b, c) Force vector map in regions 1 and 2 for P=10mW. The four-angle stars indicate the positions of zero force.

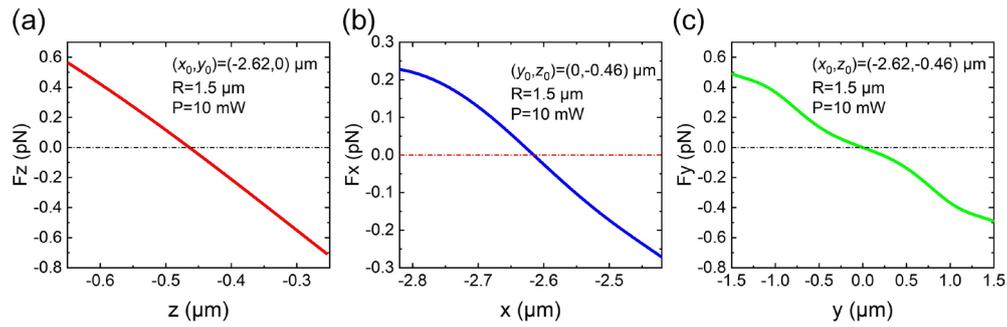

Fig. 3. The net forces experienced by a 1.5-μm-in-radius gold particle trapped by a Gaussian beam with trapping power of 10 mW in the vicinity of the position of ($x_0$, $y_0$, $z_0$) = (-2.62, 0, -0.46) μm. (a) The longitudinal forces in the z direction. (c) The transverse forces in the x direction. (d) The transverse forces in the y direction.

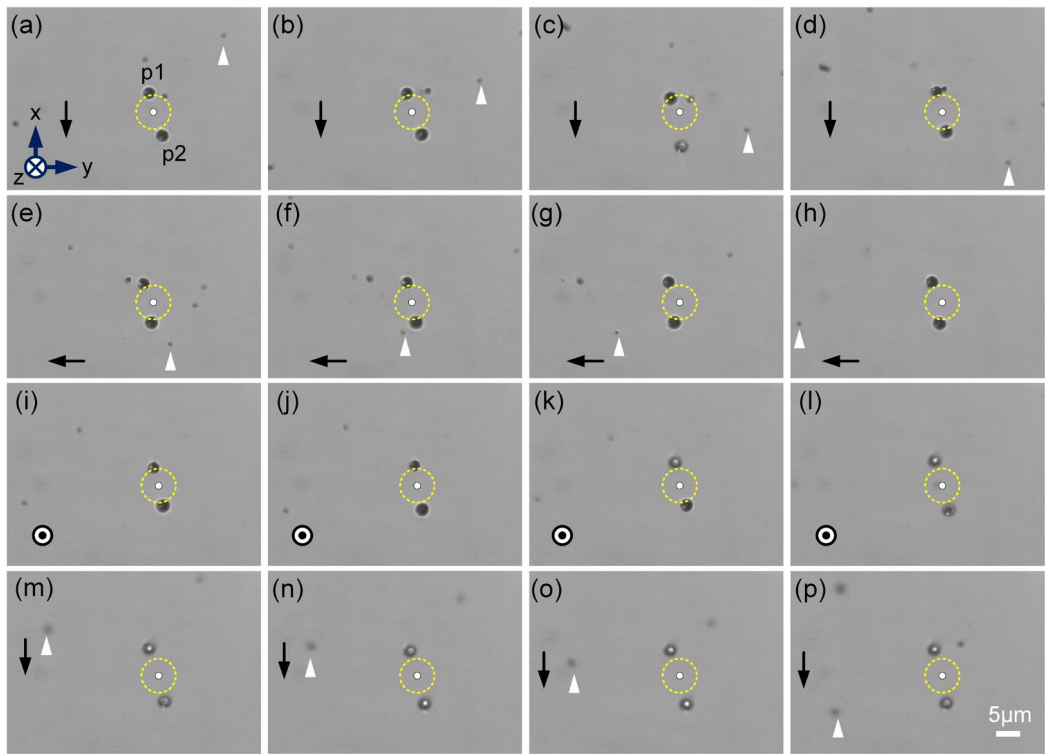

Fig. 4. Experimental results of off-axis 3D optical confinement of metallic microparticles (see Visualization 1). (a-d) The sample stage moves along the x direction. (e-h) The sample moves along the y direction. (i-l) The sample moves along the z direction. (m-p) The sample moves along the x direction after it is moved along the z direction over a distance of 10μm. The output laser power is 1W. Dark arrow: the moving direction of sample stage; white triangles: the reference objects; white dot: the position of Gaussian trap; yellow dashed circle: the possible trapping positions; p1 and p2: the positions of two hot spots.

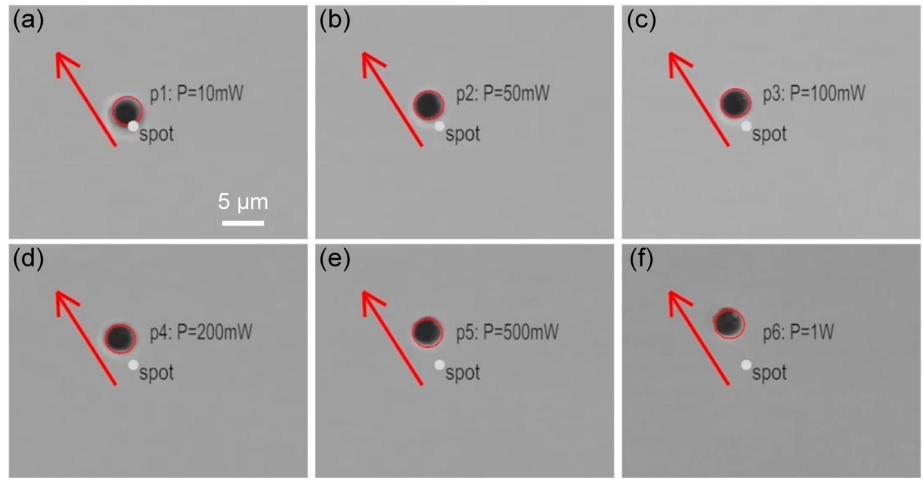

Fig. 5. Trapping position of metallic microparticle for various trapping power (see Visualization 2). (a-f) Trapping results for trapping power of 10, 50, 100, 200, 500 and 1000mW, respectively. The radius of the trapped particle is about 1.6 μm.

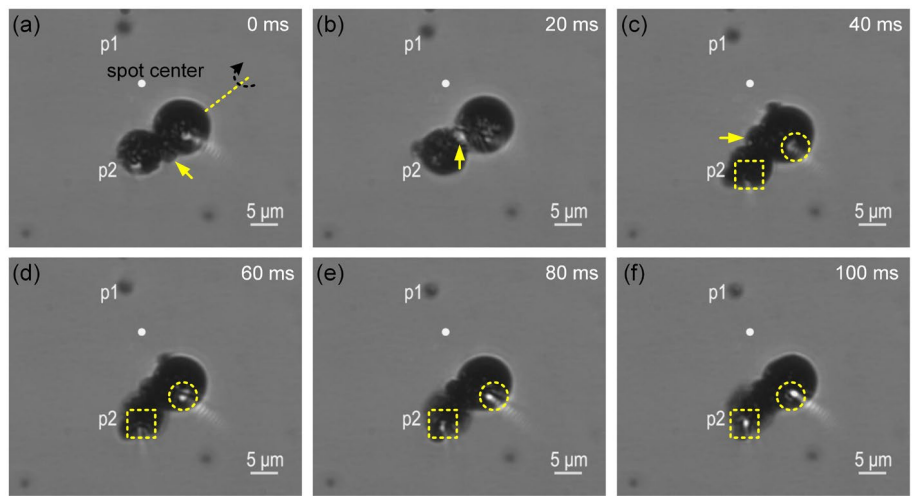

Fig. 6. Transverse spinning of metallic microparticles using Gaussian beam (see Visualization 3). (a-f) Time-lapse images of transverse spinning of metallic microparticles extracted from a video.

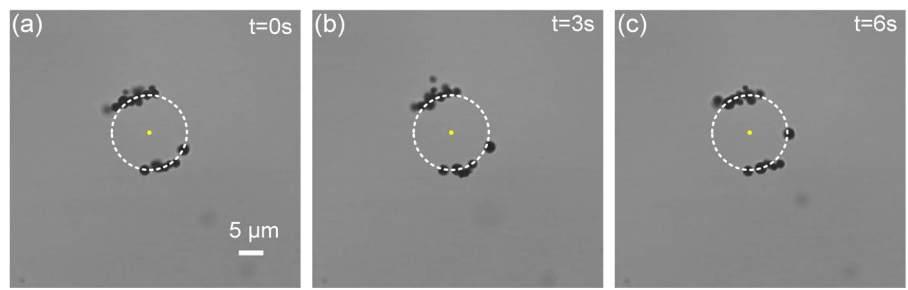

Fig. 7. Optical assembly of metallic microparticles based on off-axis optical trapping (Visualization 5). (a-c) Time-lapse images from a video. The output laser power is 1.5W.

**Funding.**

This work was supported by Natural Science Foundation of China (NSFC) (61905189); China Postdoctoral Science Foundation (2019M663656); Special guidance funds for the construction of world-class universities (disciplines) and characteristic development in Central Universities (PY3A079); National Key Research and Development Program of China (2017YFC0110100); Key Research Program of Frontier Sciences, Chinese Academy of Sciences (QYZDB-SSW-JSC005).

**References**

1. E. A. Coronado, E. R. Encina, and F. D. Stefani, "Optical properties of metallic nanoparticles: manipulating light, heat and forces at the nanoscale," Nanoscale **3**, 4042-4059 (2011).
2. J. D. Gibson, B. P. Khanal, and E. R. Zubarev, "Paclitaxel-Functionalized Gold Nanoparticles," J. Am. Chem. Soc. **129**, 11653-11661 (2007).
3. G. J. Nusz, S. M. Marinakos, A. C. Curry, A. Dahlin, F. Höök, A. Wax, and A. Chilkoti, "Label-Free Plasmonic Detection of Biomolecular Binding by a Single Gold Nanorod," Anal. Chem. **80**, 984-989 (2008).
4. S. K. Dondapati, T. K. Sau, C. Hrelescu, T. A. Klar, F. D. Stefani, and J. Feldmann, "Label-free Biosensing Based on Single Gold Nanostars as Plasmonic Transducers," Acs Nano **4**, 6318-6322 (2010).
5. T. H. Taminiau, F. D. Stefani, F. B. Segerink, and N. F. van Hulst, "Optical antennas direct single-molecule emission," Nat. Photonics **2**, 234 (2008).
6. P. Bharadwaj, B. Deutsch, and L. Novotny, "Optical Antennas," Adv. Opt. Photon. **1**, 438-483 (2009).
7. V. Kotaidis, and A. Plech, "Cavitation dynamics on the nanoscale," Appl. Phys. Lett. **87**, 213102 (2005).
8. W. Huang, W. Qian, and M. A. El-Sayed, "Gold Nanoparticles Propulsion from Surface Fueled by Absorption of Femtosecond Laser Pulse at Their Surface Plasmon Resonance," J. Am. Chem. Soc. **128**, 13330-13331 (2006).
9. L. R. Hirsch, R. J. Stafford, J. A. Bankson, S. R. Sershen, B. Rivera, R. E. Price, J. D. Hazle, N. J. Halas, and J. L. West, "Nanoshell-mediated near-infrared thermal therapy of tumors under magnetic resonance guidance," PNAS **100**, 13549-13554 (2003).
10. S. Lal, S. E. Clare, and N. J. Halas, "Nanoshell-Enabled Photothermal Cancer Therapy: Impending Clinical Impact," Acc. Chem. Res. **41**, 1842-1851 (2008).
11. K. Svoboda, and S. M. Block, "Optical trapping of metallic Rayleigh particles," Opt. Lett. **19**, 930-932 (1994).
12. Q. Zhan, "Trapping metallic Rayleigh particles with radial polarization," Opt. Express **12**, 3377-3382 (2004).
13. P. M. Hansen, V. K. Bhatia, N. Harrit, and L. Oddershede, "Expanding the Optical Trapping Range of Gold Nanoparticles," Nano Lett. **5**, 1937-1942 (2005).
14. J.-Q. Qin, X.-L. Wang, D. Jia, J. Chen, Y.-X. Fan, J. Ding, and H.-T. Wang, "FDTD approach to optical forces of tightly focused vector beams on metal particles," Opt. Express **17**, 8407-8416 (2009).
15. F. Hajizadeh, and S. N. S.Reihani, "Optimized optical trapping of gold nanoparticles," Opt. Express **18**, 551-559 (2010).
16. L. Huang, H. Guo, J. Li, L. Ling, B. Feng, and Z.-Y. Li, "Optical trapping of gold nanoparticles by cylindrical vector beam," Opt. Lett. **37**, 1694-1696 (2012).

17. L. Jauffred, S. M. Taheri, R. Schmitt, H. Linke, and L. B. Oddershede, "Optical Trapping of Gold Nanoparticles in Air," Nano Lett. **15**, 4713-4719 (2015).
18. H. Furukawa, and I. Yamaguchi, "Optical trapping of metallic particles by a fixed Gaussian beam," Opt. Lett. **23**, 216-218 (1998).
19. P. C. Ke, and M. Gu, "Characterization of trapping force on metallic Mie particles," Appl. Opt. **38**, 160-167 (1999).
20. M. Gu, D. Morrish, and P. C. Ke, "Enhancement of transverse trapping efficiency for a metallic particle using an obstructed laser beam," Appl. Phys. Lett. **77**, 34-36 (2000).
21. M. Gu, and D. Morrish, "Three-dimensional trapping of Mie metallic particles by the use of obstructed laser beams," J. Appl. Phys. **91**, 1606-1612 (2002).
22. Y. Zhang, X. Dou, Y. Dai, X. Wang, C. Min, and X. Yuan, "All-optical manipulation of micrometer-sized metallic particles," Photon. Res. **6**, 66-71 (2018).
23. K. Sakai, and S. Noda, "Optical trapping of metal particles in doughnut-shaped beam emitted by photonic-crystal laser," Electron. Lett **43**, 107-108 (2007).
24. Z. Shen, L. Su, X. C. Yuan, and Y. C. Shen, "Trapping and rotating of a metallic particle trimer with optical vortex," Appl. Phys. Lett. **109**, 241901 (2016).
25. N. B. Simpson, K. Dholakia, L. Allen, and M. J. Padgett, "Mechanical equivalence of spin and orbital angular momentum of light: an optical spanner," Opt. Lett. **22**, 52-54 (1997).
26. B. P. S. Ahluwalia, X. C. Yuan, and S. H. Tao, "Transfer of 'pure' on-axis spin angular momentum to the absorptive particle using self-imaged bottle beam optical tweezers system," Opt. Express **12**, 5172-5177 (2004).
27. P. Galajda, and P. Ormos, "Rotors produced and driven in laser tweezers with reversed direction of rotation," Appl. Phys. Lett. **80**, 4653-4655 (2002).
28. M. Padgett, and R. Bowman, "Tweezers with a twist," Nat. Photonics **5**, 343 (2011).
29. T. A. Nieminen, H. Rubinsztein-Dunlop, N. R. Heckenberg, and A. I. Bishop, "Numerical modelling of optical trapping," Comput. Phys. Commun. **142**, 468-471 (2001).
30. S. Yan, and B. Yao, "Radiation forces of a highly focused radially polarized beam on spherical particles," Phys. Rev. A **76**, 053836 (2007).
31. S. Adachi, *The Handbook on Optical Constants of Metals: In Tables and Figures* (World Scientific, 2012).